\documentclass[12pt]{iopart}
\usepackage{iopams}

\usepackage{harvard}
\usepackage{graphicx}
\usepackage{url}
\usepackage{setstack}
\usepackage[hypertexnames=false]{hyperref}
\newcommand{\nn}{\nonumber\\}
\newcommand{\gl}{\gamma_\mathrm{L}}
\newcommand{\gr}{\gamma_\mathrm{R}}
\newcommand{\pl}{\phi_\mathrm{L}}
\newcommand{\pr}{\phi_\mathrm{R}}
\newcommand{\bd}[1]{\vskip.5\baselineskip
\noindent
\textbf{#1}
\vskip.25\baselineskip
\noindent}

\usepackage{color}

\begin{document}

\title[Preprint (the version before review): Takagi et al., {\it Fluid Dyn. Res.} (2024 in press)]{Implementation of spectral methods on Ising machines: toward flow simulations on quantum annealer\footnote{This is the version of the article before peer review or editing, as submitted by an author to {\it Fluid Dynamics Research} on 19 July 2024. IOP Publishing Ltd is not responsible for any errors or omissions in this version of the manuscript or any version derived from it. The Accepted Manuscript after revision is available online at \url{https://doi.org/10.1088/1873-7005/ad8d09}}}

\author{Kenichiro~Takagi$^1$, 
Naoki~Moriya$^1$, 
Shiori~Aoki$^1$, 
Katsuhiro~Endo$^2$, 
Mayu~Muramatsu$^1$, 
and Koji~Fukagata$^1$}

\address{
$^1$~Department of Mechanical Engineering, Keio University, Yokohama, 223-8522, Japan \\
$^2$~National Institute of Advanced Industrial Science and Technology, Tsukuba, 305-8568, Japan
}
\ead{fukagata@mech.keio.ac.jp}


\begin{abstract}
We investigate the possibility and current limitations of flow computations using quantum annealers by 
solving
a fundamental flow problem on Ising machines.
As a fundamental problem, we consider
the one-dimensional advection-diffusion equation. 
We formulate it in a form suited to Ising machines (i.e., both classical and quantum annealers),
perform extensive numerical tests on a classical annealer,
and finally test 
it on an actual quantum annealer.
To make it possible to process with an Ising machine, the problem is formulated as a minimization problem of the residual of the governing equation discretized using either the spectral method or the finite difference method.
The resulting system equation is then converted to the Quadratic Unconstrained Binary Optimization (QUBO)
form though quantization of variables.
It is found in the numerical tests using a classical annealer that the spectral method requiring smaller number of variables 
has a particular merit over the finite difference method
because the accuracy deteriorates with the increase of the number of variables.
We also found that the 
computational error
varies depending on the condition number of the coefficient matrix.
In addition, we extended it to a two-dimensional problem
and confirmed its fundamental applicability.
From the numerical test using a quantum annealer, however, it turns out that the computation 
using a quantum annealer is
still challenging due largely to the structural difference from the classical annealer, which leaves a number of issues toward its practical use.

\end{abstract}

\vspace{2pc}


\section{Introduction}\label{s:Intro}
Over the past 40 years, Direct Numerical Simulation (DNS) has played a key role in improving our fundamental understanding of fluid flow phenomena (Rogallo \harvardand\ Moin 1984, Kim \harvardand\ Leonard 2024).
Nowadays, DNS has become a inevitable tool not only for elucidating detailed flow physics 
(Moin \harvardand\ Mahesh 1998, Alfonsi 2011)
but also for predicting the performance of flow control methods at a thought experiment level (Kim 2003, Ricco {\it et al} 2021, Fukagata {\it et al} 2024).
However, the number of grid points required in DNS amounts to the order of $\mathrm{Re}^{9/4}$,
making it
impractical to use DNS for industrial problems.
Although extensive efforts have also been made to increase the generality of closure models for Reynolds-Averaged Navier--Stokes (RANS) and Large Eddy Simulation (LES), developing such general closure models remains very challenging (Argyropoulos \harvardand\ Markatos 2015).

Recently, use of
quantum computers 
has attracted increasing attentions in various research fields (Sood \harvardand\ Pooja 2023).
Quantum computers are expected to be faster than classical computers by taking advantage of phenomena unique to quantum mechanics --- superposition, branching, interference, and entanglement.
While a classical bit can only take one of two states, 0 or 1, a qubit (i.e., a quantum bit) can take a superposition of both 0 and 1 states.
Since the entire system of the quantum computer has exponentially overlapping states, a large number of operations can be performed in parallel in an exponential 
manner.
The branching and interference effects of the qubits allow each computation to interact with the other rather than being isolated.
The result of the quantum computation can be observed as a single result of many overlapping calculations, and the only desired solution is emphasized by quantum interference.
Thus, parallelism can be used more efficiently by taking advantage of the quantum superposition principle.
Furthermore, due to quantum mechanical correlations called quantum entanglement, qubits are strongly connected each other independent of distance.
The action of quantum entanglement enables exponential parallelism which is unable to explain by the classical theory.
Based on these features of quantum computation, quantum computers are expected to perform parallel computations with incomparably greater efficiency than classical computers~\cite{G1999}.

There are two main quantum computer architectures that have been proposed.
The first is called a gate-based quantum computer (Barenco {\it et al} 1995).
This
type of quantum computer
 performs calculations by stacking quantum logic gates that resemble the logic gates of the classical computer.
Therefore,
implementation 
of an existing computational algorithm
on gate-based quantum computers is expected to be relatively easy.
For a fluid flow problem,
Gaitan (2020)
applied the quantum algorithm for solving nonlinear ordinary differential equations proposed by 
Kacewicz (2006)
to simulate a Laval nozzle including shock wave capture using gate-based quantum computers.
More recently,
Gourianov {\it et al} (2022)
proposed a method for computing turbulent structures
based on a matrix product state, which is considered 
suited to computations
on quantum gates by effectively using quantum entanglement \cite{F2022}.
On the other hand, quantum gates are susceptible to noise and errors, and require a large number of error-correcting bits.
For this reason, 
fluid computations using gate-based quantum computers
are still far from practice.

The second type
is called a quantum annealer (Santoro {\it et al} 2002).
Quantum annealing is a type of quantum adiabatic computation that is specialized for solving combinatorial optimization problems.
In a quantum annealer, all qubits involved in a computation are coupled and remain in the ground state with the lowest overall energy.
Therefore, the entire annealing system is affected by noise, resulting in relatively less noise effect than that of gate-based system.
For this reason, the quantum annealer has been put into practical use prior to quantum gates, and is now adopted by major research institutes and companies in Spain, the United States, and Japan~\cite{DWa2023}.
The combinatorial optimization problems solved by quantum annealers
are called the Ising 
model,
and they 
are NP-hard.
The computer containing the quantum annealer that performs the computation of the Ising model
(i.e., the equivalent Quadratic Unconstrained Binary Optimizations (QUBOs))
 is called an Ising machine.
No algorithm has been found that can solve NP-hard problems in polynomial time using conventional computers, and exponential time is required to find the optimal solution. 
However, quantum annealing is expected to make it possible to solve the problem in polynomial time.
Also, since the Ising model is NP-hard, many NP-complete problems and some NP-hard problems can be efficiently embedded in the Ising model and solved~\cite{L2014}.

However, 
very limited studies have been reported so far on the
applications 
of quantum annealers
to fluid dynamics 
problems, 
since quantum annealing has developed primarily to focus on combinatorial optimization problems.
Ray 
{\it et al} (2022) attempted to
solve the 
laminar Poiseuille
flow 
on
the D-Wave
quantum annealer.
They discretized the one-dimensional governing equation by the finite difference method (FDM) and formulated the problem as a least-square minimization of the residual, 
whereby the governing equation
can be converted to QUBO form.
However, their simulation results revealed that
significant challenges remain in computational accuracy.
On classical computers,
the computational accuracy 
can be improved by increasing the number of 
computational
points.
On the other hand, calculations using the quantum annealer are essentially combinatorial optimizations, and the accuracy of the obtained solution tends to decrease as the number of qubits increases.
Therefore, 
solutions with 
a reasonable accuracy
could not be 
obtained
for 
such a
simple 
problem of one-dimensional laminar Poiseuille flow.

As a different approach,
Kuya 
{\it et al} (2024)
implemented the lattice gas method on the Ising machine and 
applied it to a laminar Poiseuille flow.
This method was suggested to 
have a high affinity
with quantum annealing because the presence or absence of particles at each lattice point can be expressed as 0 or 1.
Their simulation was able to 
accurately
represent 
the
collisions 
between particles.
However, the lattice gas method has a problem that the computational variables increase significantly due to megascaling, and a considerably larger machine is required to perform 
computations
 using the actual quantum annealer, which is currently not available.

As 
introduced above, the 
computational
accuracy 
on a
quantum annealer
decreases as the size of the problem and the number of variables increase.
Therefore, a scheme 
requiring a smaller number of variables is 
preferred, and
it is straightforward to examine the possibility of the spectral method --- a mature method in DNS on classical computers --- in addition to FDM studied by Ray {\it et al} (2022).

In the present study, we attempt to use the spectral method
to reduce the number of required variables on a quantum annealer.
As fundamental problems, we consider
the 
steady and unsteady
advection-diffusion equations.
We formulate the governing equations in the
QUBO 
form
and 
perform computations
on the Ising machine.
The primary motivation of the present work is to reveal the current possibilities and limitations of flow computations on a quantum annealer.
For the next 40 years of computational fluid dynamics,
here we start this investigation from a na\"ive implementation similar to Ray {\it et al} (2022).
The paper is organized as follows:
In \Sref{s:Methods}, we explain the basic terms and knowledge of Ising machines and quantum annealing.
Then, we describe how to attribute the partial differential equations using the spectral method to QUBO.
In \Sref{s:Ne}, we perform numerical experiments using 
the
Ising model 
on
a classical computer, 
i.e.,
simulated annealing.
In \Sref{s:IiQA}, we perform 
computations
on the actual quantum annealer and compare them with simulated annealing.
Finally,
summary and 
outlook
are provided in \Sref{s:Conc}. 

\section{Methods}\label{s:Methods}
\subsection{Quantum annealing}\label{s:QA}
Quantum annealing is a computational scheme based on the solution of an NP-hard combinatorial optimization problem called the transverse field Ising model~\cite{L2014}.
The transverse field Ising model is a mathematical model of ferromagnetism in statistical mechanics.
In the model, atoms are located at lattice points (sites) in a crystal, and each electron is considered to have either upward or downward spin.
Since the electron spins form a magnetic field, the crystal is considered to be magnetized if the spins on the sites are oriented uniformly, and unmagnetized if the spins are oriented randomly
(Santoro {\it et al} 2002).
The directions of these spins 
are
determined by the directions of spins in other surrounding sites 
and 
the 
external magnetic field, so that the overall energy is minimized.
There are $2^\nu$ possible spin configurations for a system with $\nu$ sites, but no algorithm has been found to calculate the solution with the lowest energy in polynomial time.
Therefore, when using the classical computer, the only way to determine the solution is to heuristically search from an exponentially large number of value ranges.
In addition, since a full search is not possible with the classical computer, there is always a risk of falling into a local solution.
Quantum annealing is the computation scheme that can be used to compute the transverse field Ising model at high speed.
In quantum annealing, the Ising model is reproduced by connecting qubits to each other, and the global energy minimum solution is searched for by initially increasing the fluctuation of the state to promote state transitions and gradually decreasing the fluctuation of the system state to avoid falling into a localized solution~\cite{KN1998}.

The Hamiltonian $H(t)$, which represents the state of the transverse magnetic field Ising model, is given as,
\begin{eqnarray}
    H&=-\sum_{i<j}J_{ij}\sigma_i^z\sigma_j^z-\sum_{i}h_i\sigma_i^z-\Gamma(t)\sum_i\sigma_i^x\label{eq:tmIm},\\
    &=H_0-\Gamma(t)\sum_i\sigma_i^x,
\end{eqnarray}
where $J_{ij}$ 
denotes
the interaction between site $i$ and site $j$, $h_i$ is the external magnetic field applied to site $i$, and $\sigma_i^x$ and $\sigma_i^z$ are the longitudinal and transverse Pauli matrices acting on site $i$, respectively.
The function $\Gamma(t)$ is 
set
sufficiently large in the initial state $t=0$ and decreases to zero as time $t$ increases.
The first and second terms in \Eref{eq:tmIm} are collectively denoted as $H_0$.
The Pauli matrix
$\sigma^x$ in \Eref{eq:tmIm} acts to invert the spin direction at each site, and $\sigma^z$ to discriminate the spin direction, respectively.
Thus, the initial state of the Hamiltonian is dominated by $\sigma^x$, which leads to large fluctuations of the state.
The spin of each site is undetermined, and the system tends to approach the ground state with the lowest energy.
The energy of the final state is given as,
\begin{equation}
    E(s_1,s_2,\cdots)=-\sum_{i<j}J_{ij}s_is_j-\sum_{i}h_is_i,\ s_i\in \{-1,+1\},\label{eq:Isings}
\end{equation}
where $s_i$ represents the spin direction on the site $i$.
If the spin at the site $i$ is upward, $s_i=+1$; if it is downward, $s_i=-1$.

Since the spin configuration obtained by the calculation of the transverse field Ising model is determined by the effects of the interaction $J_{ij}$ and the local magnetic field $h_i$, the calculation can be applied to other problems by setting the values of $J_{ij}$ and $h_i$.
\Eref{eq:Isings} is equivalent to the quadratic form of the binary array,
\begin{equation}
    f(q_1,q_2,\cdots)=\sum_{i,j}Q_{ij}q_iq_j,\ q_i\in\{0,1\}.\label{eq:f_QUBO}
\end{equation}
where $q_i$ and $Q_{ij}$ are given by
\begin{eqnarray}
    q_i&=\frac{1+s_i}{2}\label{eq:Ising_to_QUBO;v},\\
    Q_{ij}&=2J_{ij}+\left[h_i-\sum_k\left(J_{ik}+J_{ki}\right)\right]\delta_{ij}\label{eq:Ising_to_QUBO;Q}.
\end{eqnarray}
\Eref{eq:f_QUBO} is the objective function of Quadratic Unconstrained Binary Optimization (QUBO), i.e., the function that the quantum annealers are designed to minimize.
Since the value range of the variables is $\{0,1\}$, it matches the value range of the classical bits and is easier to apply than the transverse field Ising model.\\

\subsection{Linear least squares method by quantum annealer}\label{s:LlsmQA}
Partial differential equations can be reduced to a set of linear equations 
through 
discretization.
In order to make this 
computable with quantum annealing, the 
equation
is 
translated into the
QUBO 
form
and the continuous quantities are represented by a linear sum of qubits.

The set of linear equations can be expressed as
\begin{eqnarray}
    \boldsymbol{Ax}-\boldsymbol{b}=\boldsymbol{0}\label{eq:LE},
\end{eqnarray}
where $\boldsymbol{A}$ $\in \mathbb{R}^{N\times N}$ is a constant matrix, $\boldsymbol{x}$ $=(x_1, x_2,\cdots, x_N)^T$ is the design variable, and $\boldsymbol{b}$ $\in \mathbb{R}^{N}$ is a constant vector.
The linear least squares method solves \Eref{eq:LE} as a minimization problem of the form,
\begin{eqnarray}
    \boldsymbol{x}&=\underset{\boldsymbol{x}}{\mathrm{argmin}}\left|\boldsymbol{Ax}-\boldsymbol{b} \right|^2,\nn
    &=\underset{\boldsymbol{x}}{\mathrm{argmin}}\left(\boldsymbol{x}^\mathrm{T}\boldsymbol{A}^\mathrm{T}\boldsymbol{Ax}-2\boldsymbol{b}^\mathrm{T}\boldsymbol{Ax}+\mathrm{const.}\right),
\end{eqnarray}
where $\boldsymbol{A}^\mathrm{T}$ is a transposed matrix of $\boldsymbol{A}$.
Let $f$ be 
the
objective function,
$f$ reads
\begin{eqnarray}
    f&=\left|\boldsymbol{Ax}-\boldsymbol{b}\right|^2\nn
    &=\boldsymbol{x}^\mathrm{T}\boldsymbol{A}^\mathrm{T}\boldsymbol{Ax}-2\boldsymbol{b}^\mathrm{T}\boldsymbol{Ax}+\mathrm{const.}\label{eq:objfunc_LR}
\end{eqnarray}
If we consider quantization 
of the continuous design variables $x_i \in \left[x_\mathrm{min},x_\mathrm{max}\right]$
with $n$ qubits $q_1, q_2,\cdots q_n$, 
it
can be expressed as 
\begin{eqnarray}
    x_i&=x_\mathrm{min}+\frac{x_\mathrm{max}-x_\mathrm{min}}{1-2^{-n}}\sum_{k=1}^{n}2^{-k}q_{(i-1)n+k}\nn
    &\equiv x_\mathrm{min}+\sum_{k=1}^{n}\epsilon_kq^i_k
    \label{eq:exp_of_x},
\end{eqnarray}
where 
\begin{equation}
    \epsilon_k=\frac{2^{-k}(x_\mathrm{max}-x_\mathrm{min})}{1-2^{-n}}
\end{equation}
and
\begin{equation}
    q^i_k=q_{(i-1)n+k}.
\end{equation}
Note that this quantization is similar to that used in Ray {\it et al} (2022) but with a normalization based the upper and lower bounds, $x_\mathrm{max}$ and $x_\mathrm{min}$.

Substituting 
these quantized variables
into \Eref{eq:objfunc_LR} yields
\begin{eqnarray}
    \fl f(q_1,q_2,\cdots)&=\sum_{h,i,j}A_{hi}A_{hj}x_ix_j-2\sum_{i,j}b_jA_{ji}x_i+\sum_ib_ib_i,\nn
    &=\sum_{h,i,j,k,l}A_{hi}A_{hj}\epsilon_k\epsilon_lq^i_kq^j_l\nn
    &\ \ \ +\sum_{i,k}\left[\sum_{h,s}(A_{ih}A_{hs}+A_{sh}A_{hi})x_\mathrm{min}-2\sum_{h}b_jA_{hi}\right]\epsilon_kq^i_k+\mathrm{const.}\label{eq:nQUBO}
\end{eqnarray}
Since $q_iq_i=q_i$, $f$ 
can be written as
\begin{eqnarray}
    \fl f(q_1,q_2,\cdots)\nn
    \fl =\sum_{i,j,k,l}\left\{\sum_{h}A_{hi}A_{hj}\epsilon_k\epsilon_l+\left[\sum_{h,s}(A_{ih}A_{hs}+A_{sh}A_{hi})x_\mathrm{min}-2\sum_{h}b_hA_{hi}\right]\delta_{ij}\delta_{kl}\epsilon_k\right\}q^i_kq^j_l,\label{eq:QUBOex}
\end{eqnarray}
where $\delta_{ij}$ is the Kronecker delta.
Since \Eref{eq:QUBOex} is a quadratic form of $q_i$ of the same form as \Eref{eq:f_QUBO}, \Eref{eq:QUBOex} can be used as the objective function of QUBO and can be handled by quantum annealing.

\subsection{Discretization of matrices}
We consider advection-diffusion equations as example problems in the present study.
For instance, the one-dimensional advection-diffusion equation is written as
\begin{equation}
    \frac{\partial \phi}{\partial t}=-u\frac{\partial \phi}{\partial x}+\alpha\frac{\partial^2 \phi}{\partial x^2},\quad (0\leq x\leq 1),\label{eq:1DADeq}
\end{equation}
where $\phi$ is a passive scalar, $u$ is a flow velocity, and $\alpha$ is a diffusion coefficient.
The boundary conditions are $\phi=\pl$ at the left end $x=0$ and $\phi=\pr$ at the right end $x=1$.
\Eref{eq:1DADeq} is discretized 
using
the Finite Difference Method (FDM) or the Chebyshev Spectral Method (CSM)
as detailed below.
In both cases, once the objective function $f$
is written using $\boldsymbol{A}$ and $\boldsymbol{b}$, it can be converted to the QUBO form using the
quantization \eref{eq:exp_of_x}.

\subsubsection{Finite Difference Method (FDM)}
\ 

\noindent
For explanatory purpose, here we assume the
number of 
the computational
points 
to be $M=4$ except for the end points.
In this case, the spatial
grid
 width $\Delta x$ is $\Delta x=1/5$.
Discretizing the 
time integration 
using the Euler implicit method, \Eref{eq:1DADeq} can be expressed in the form of \Eref{eq:LE}, i.e., $\boldsymbol{A}\boldsymbol{\phi}-\boldsymbol{b}=\boldsymbol{0}$, with
\begin{eqnarray}
    \boldsymbol{A}&=\left[\matrix{%
        1+2D & \frac{1}{2}C-D & 0 & 0 \cr
        -\frac{1}{2}C-D & 1+2D & \frac{1}{2}C-D & 0 \cr
        0 & -\frac{1}{2}C-D & 1+2D & \frac{1}{2}C-D \cr
        0 & 0 & -\frac{1}{2}C-D & 1+2D \cr
    }\right],\\
    \boldsymbol{b}&=\left[\matrix{%
        \phi^{\imath}_1-\left(-\frac{1}{2}C-D\right)\pl \cr
        \phi^{\imath}_2 \cr
        \phi^{\imath}_3 \cr
        \phi^{\imath}_4-\left(\frac{1}{2}C-D\right)\pr
    }\right],
\end{eqnarray}
and,
\begin{equation}
    \boldsymbol{\phi}=
        \left[\matrix{%
        \phi^{\imath+1}_1 \cr
        \phi^{\imath+1}_2 \cr
        \phi^{\imath+1}_3 \cr
        \phi^{\imath+1}_4 \cr
    }\right],
\end{equation}
with $\Delta t$ as the time step width, where $C$ is the signed Courant number $C=u\Delta t/\Delta x$, $D$ is the diffusion number $D=\alpha\Delta t/(\Delta x)^2$, and $\phi_n^\imath$ is the value of $\phi$ at the $n$-th grid point $x = n\Delta x$ at time $\imath\Delta t$.
In this case, the objective function is defined by \Eref{eq:objfunc_LR} as it is.

\subsubsection{Chebyshev Spectral Method (CSM)}\label{s:CSM}
\ 

\noindent
The CSM is a spectral method often used 
when the computational domain has two ends, 
e.g., DNS of turbulent channel flows.
In the CSM, the function is expanded in terms of Chebyshev polynomials of the first kind (hereafter simply Chebyshev polynomials), one of the orthogonal polynomials.
The $n$th-order Chebyshev polynomial $T_k(\xi)$ is defined as,
\begin{equation}
    T_k(\xi)=\cos{(n\arccos{\xi})},\label{eq:def_o_T_k}
\end{equation}
for $-1\leq\xi\leq1$.
This polynomial exhibits orthogonality, i.e.,
\begin{eqnarray}
    \int_{-1}^{+1}T_i(\xi)T_j(\xi)\frac{\mathrm{d}\xi}{\sqrt{1-\xi^2}}=
    \left\{\matrix{%
        \pi, & i=j=0 \cr
        \pi/2, & i=j\neq 0 \cr
        0, & i\neq j
    }\right.,\label{eq:orth_o_T}
\end{eqnarray}
when multiplied by a weight function $(1-\xi^2)^{-1/2}$.
The derivative of this polynomial is given by a linear sum of polynomials of lower degree; thus,
\begin{eqnarray}
    \frac{\mathrm{d}}{\mathrm{d}\xi}T_k(\xi)=\left\{\matrix{%
        2k\sum_{1\leq i \leq k/2}T_{2i-1}(\xi) & k:\mathrm{even}, \cr
        2k\left[\sum_{1\leq i \leq (k-1)/2}T_{2i}(\xi)+\frac{1}{2}T_0(\xi)\right] & k:\mathrm{odd}. \cr
    }\right.\label{eq:dxi_o_T}
\end{eqnarray}

We 
discretize
\Eref{eq:1DADeq} 
by the CSM
to form a minimization problem in \Sref{s:LlsmQA}.
Since
the domain of \Eref{eq:1DADeq} is $0\leq x \leq 1$, 
a variable transformation $\xi = 2x-1$ is first applied, such that
\begin{eqnarray}
    \frac{\partial \phi}{\partial t}=-2u\frac{\partial \phi}{\partial \xi}+4\alpha\frac{\partial^2 \phi}{\partial \xi^2},\ -1\leq \xi \leq +1\label{eq:1DADxi}.
\end{eqnarray}
Then, by substituting the Chebyshev expansion of
$\phi$, i.e.,
\begin{eqnarray}
    \phi(t,\xi) = \sum_{k=0}^N a_k(t)T_k(\xi),\label{eq:exp_phi}
\end{eqnarray}
into \Eref{eq:1DADxi}, the residual $R(t,\xi)$ of the equation is expressed as,
\begin{eqnarray}
    R(t,\xi)&=\sum_k r_k(t) T_k(\xi)\nn
    &=\sum_k \left\{\frac{\mathrm{d}a_k(t)}{\mathrm{d}t}T_k(\xi)+2ua_k(t)\frac{\mathrm{d}T_k(\xi)}{\mathrm{d}\xi}-4\alpha a_k(t)\frac{\mathrm{d}^2T_k(\xi)}{\mathrm{d}\xi^2}\right\},\label{eq:def_o_R}
\end{eqnarray}
where $N$ is the expansion degree.
By discretizing this
in 
time 
as well, we obtain an expression relating the Chebyshev spectrum $a$ of the physical quantity $\phi$ to the Chebyshev spectrum $r$ of the residual $R$.
For example, if we 
take
$N=3$ and use the Euler implicit method for time integration
--- again for the explanatory purpose ---, \Eref{eq:def_o_R} can be written as
\begin{eqnarray}
    R(t,\xi)&=\sum_{k=0}^3 r_k(t)T_k(\xi)\nn
    &=\left\{(\Delta t)^{-1}a_0^{\imath+1}+2ua_1^{\imath+1}-16\alpha a_2^{\imath+1}+6ua_3^{\imath+1}-(\Delta t)^{-1}a_0^{\imath}\right\}T_0(\xi)\nn
    &+\left\{(\Delta t)^{-1}a_1^{\imath+1}+8ua_2^{\imath+1}-96\alpha a_3^{\imath+1}-(\Delta t)^{-1}a_1^{\imath}\right\}T_1(\xi)\nn
    &+\left\{(\Delta t)^{-1}a_2^{\imath+1}+12ua_3^{\imath+1}-(\Delta t)^{-1}a_2^{\imath}\right\}T_2(\xi)\nn
    &+\left\{(\Delta t)^{-1}a_3^{\imath+1}-(\Delta t)^{-1}a_3^{\imath}\right\}T_3(\xi)\label{eq:Res}
\end{eqnarray}
where $a^{\imath}$ and $a^{\imath+1}$ are the $n$th-order spectra at time $\imath\Delta t$ and $(\imath+1)\Delta t$, respectively.
To have $R=0$,
the condition imposed on $a$ is in the form of 
\Eref{eq:LE}, i.e., $\boldsymbol{A}\boldsymbol{a}-\boldsymbol{b}=\boldsymbol{0}$ with
\begin{equation}
    \boldsymbol{A}=
    \left[\matrix{%
        (\Delta t)^{-1} & 2u & -16\alpha & 6u \cr
        0 & (\Delta t)^{-1} & 8u & -96\alpha \cr
        0 & 0 & (\Delta t)^{-1} & 12u \cr
        0 & 0 & 0 & (\Delta t)^{-1} \cr
    }\right], \quad
    \boldsymbol{b}=
    \left[\matrix{%
        (\Delta t)^{-1}a_0^{\imath} \cr
        (\Delta t)^{-1}a_1^{\imath} \cr
        (\Delta t)^{-1}a_2^{\imath} \cr
        (\Delta t)^{-1}a_3^{\imath} \cr
    }\right]
    \label{eq:ChebAb}
\end{equation}
and
\begin{equation}
    \boldsymbol{a}=
    \left[\matrix{%
        a_0^{\imath+1} \cr a_1^{\imath+1} \cr a_2^{\imath+1} \cr a_3^{\imath+1} \cr
    }\right].
\end{equation}

As in \Sref{s:LlsmQA}, the boundary conditions are given as $\phi=\pl$ at the leftmost $x=0$ and $\phi=\pr$ at the rightmost $x=1$.
From the definition of the Chebyshev polynomial \eref{eq:def_o_T_k}, the value at the boundary is given as
\begin{eqnarray}
    T_k(\xi=-1)=(-1)^k,\ T_k(\xi=+1)=1\label{eq:BC_o_T_k},
\end{eqnarray}
and by substituting $\xi=\pm 1$ for $\phi(\xi)$, the boundary conditions are expressed as
\begin{equation}
\boldsymbol{A}_\gamma\boldsymbol{a}-\boldsymbol{b}_\gamma=\boldsymbol{0},\label{eq:BC}
\end{equation}
where
\begin{equation}
    \boldsymbol{A}_\gamma=
    \left[\matrix{%
            +\gl & -\gl & +\gl & -\gl \cr
            +\gr & +\gr & +\gr & +\gr \cr
    }\right]
    \boldsymbol{b}_\gamma=
    \left[\matrix{%
        \gl\pl \cr
        \gr\pr \cr
    }\right].
\end{equation} 
In \Eref{eq:BC}, the first row is multiplied by $\gl$ and the second row is multiplied by $\gr$.
These constants
$\gl$ and $\gr$ are parameters 
represent the weights of the boundary conditions
when the set of equations are solved.
From \Eref{eq:Res} and \Eref{eq:BC}, $a^{\imath+1}$ must satisfy,
\begin{equation}
    \left[\matrix{%
        \boldsymbol{A} \cr
        \boldsymbol{A}_\gamma \cr
    }\right]
    \boldsymbol{a}
    -
    \left[\matrix{%
        \boldsymbol{b} \cr
        \boldsymbol{b}_\gamma \cr
    }\right]
    =
    \boldsymbol{0}.\label{eq:OD}
\end{equation}

\Eref{eq:OD} cannot be solved by the direct method because there are four variables $a$ and six conditions imposed on $a$, thus creating an overdetermined system.
By solving \Eref{eq:OD} using the linear least squares method, we obtain a plausible solution $\boldsymbol{a}=\boldsymbol{A}^+\boldsymbol{b}$, where $\boldsymbol{A}^+$ is the Moore--Penrose inverse of matrix $\boldsymbol{A}$.
\Tref{tab:ex_o_OD} shows the least squares solution $\boldsymbol{a}=\boldsymbol{A}^+\boldsymbol{b}$ for various values of $\gl$ and $\gr$.
As can be seen from \tref{tab:ex_o_OD}, the least squares solution varies depending on $\gl$ and $\gr$.
Since $\gl$ and $\gr$ are arbitrary constants that have nothing to do with physical phenomena, the solution must be independent from $\gl$ and $\gr$.
Therefore, reducing the condition for the residuals \eref{eq:Res}, by the number of boundary conditions \eref{eq:BC} is necessary.
Hence, rather than strictly having $R=0$, the constraint for the residual is that each spectrum is orthogonal to certain weight functions $\psi_k$,
\begin{eqnarray}
    \int_{-1}^{+1}R(t,\xi)\psi_k(\xi)\frac{\mathrm{d}\xi}{\sqrt{1-\xi^2}}=0.\label{eq:constraint}
\end{eqnarray}
Depending on choice of the weight function $\psi_k$, 
the following three formulations are 
widely
used, i.e., Galerkin, tau, and collocation methods (Fornberg 1998).
See, Appendix A for more details on these methods.
For instance, when the tau method is adopted, \Eref{eq:OD} reduces to read
\begin{eqnarray}
    \left[\matrix{%
            (\Delta t)^{-1} & 2u & -16\alpha & 6u \cr
      		0 & (\Delta t)^{-1} & 8u & -96\alpha \cr
            +\gl & -\gl & +\gl & -\gl \cr
            +\gr & +\gr & +\gr & +\gr \cr
    }\right]    \boldsymbol{a}
    -
    \left[\matrix{%
            (\Delta t)^{-1}a_0^{\imath} \cr
        (\Delta t)^{-1}a_1^{\imath} \cr
        \gl\pl \cr
        \gr\pr \cr
    }\right]
    = \boldsymbol{0},
\end{eqnarray} 
where the first two rows are the reduced $\boldsymbol{A}$ and $\boldsymbol{b}$ derived in Appendix A.


\begin{table}[t]
    \caption{Comparison of the change in the steady state solution of the overdetermined system when varying $\gl,\gr$ and the Chebyshev coefficients for the exact steady state solution using the Discrete Fourier Transform (DFT).}
    \label{tab:ex_o_OD}
    \begin{indented}
        \item[]\begin{tabular}{@{}lllll}
        \br
            $\gl,\gr$ & $a_0$ & $a_1$ & $a_2$ & $a_3$ \\
        \mr
            $10^0$ & $0.436950$ & $0.490867$ & $0.063050$ & $0.005172$ \\
            $10^1$ & $0.436451$ & $0.494748$ & $0.063549$ & $0.005213$ \\
            $10^2$ & $0.436446$ & $0.494787$ & $0.063554$ & $0.005213$ \\
        \mr
            DFT & $0.438454$ & $0.494908$ & $0.061229$ & $0.005076$ \\
        \br
        \end{tabular}
    \end{indented}
\end{table}

In addition to the three standard methods mentioned above, we also examine another method to define the residual $R$ itself as the weight function.
Here we consider again the $N=3$ case for the explanatory purpose.
When the residual $R$ is constrained to be orthogonal to its zeroth and first order terms,
the objective function $f_\mathrm{R}$ can be expressed as
\begin{eqnarray}
    f_\mathrm{R}&=\sum_{k=0}^{N-2}\int_{-1}^{+1}R(t,\xi)r_k(t)T_k(\xi)\mathrm{d}\xi\nn
    &=
    \left[\matrix{%
        r_0 & r_1
    }\right]
    \left[\matrix{%
        I_{00} & I_{01} & I_{02} & I_{03} \cr
        I_{10} & I_{11} & I_{12} & I_{13} \cr
    }\right]
    \left[\matrix{%
        r_0 \cr
        r_1 \cr
        r_2 \cr
        r_3 \cr
    }\right]\nn
    &=\boldsymbol{a}^\mathrm{T}\boldsymbol{A}_\mathrm{R}\boldsymbol{IAa}-\left(\boldsymbol{b}_\mathrm{R}^\mathrm{T}\boldsymbol{IA}+\boldsymbol{b}^\mathrm{T}\boldsymbol{I}^\mathrm{T}\boldsymbol{A}_\mathrm{R}\right)\boldsymbol{a}
    +\mathrm{const.},\label{eq:Newf}
\end{eqnarray}
where
\begin{eqnarray}
    \fl
    \boldsymbol{A}_\mathrm{R}=
    \left[\matrix{%
        (\Delta t)^{-1} & 2u & -16\alpha & 6u \cr
        0 & (\Delta t)^{-1} & 8u & -96\alpha \cr
    }\right],\ 
    I_{ij}=\int_{-1}^{+1}T_i(\xi)T_j(\xi)\mathrm{d}\xi,\ 
    \boldsymbol{b}_\mathrm{R}=
    \left[\matrix{%
        (\Delta t)^{-1}a_0^{\imath} \cr
        (\Delta t)^{-1}a_1^{\imath} \cr
    }\right].\nonumber
\end{eqnarray}
with
$\boldsymbol{A}_\mathrm{R}$ and $\boldsymbol{b}_\mathrm{R}$ 
being
the matrices $\boldsymbol{A}$ and $\boldsymbol{b}$ in \Eref{eq:ChebAb} with the bottom 2 rows removed, respectively.
By deleting the rows corresponding to the number of boundary conditions, the system is prevented from becoming the overdetermined system.
However, 
$f_\mathrm{R}$ does not contain any information on the boundary conditions.
Therefore, we add the objective function for the boundary condition, i.e.,
\begin{eqnarray}
    f_\mathrm{BC}=
    \left|
    \boldsymbol{A}_\gamma\boldsymbol{a}-\boldsymbol{b}_\gamma
    \right|^2.\label{eq:penNewf}
\end{eqnarray}
to have the final form of the objective function, i.e.,
\begin{equation}
    f=f_\mathrm{R}+f_\mathrm{BC},
    \label{eq:fCSM}
\end{equation}
so that
a solution 
minimizing $f_\mathrm{R}$ while respecting the boundary conditions
can be obtained.
Hereafter, this fourth method for imposing the boundary conditions is referred to as the penalty method.

\subsection{Combination of Fourier and Chebyshev spectral methods}\label{s:FSM}
An extension of the present method to a two-dimensional problem 
with a periodicity is straightforward.
As an example, we consider the two-dimensional advection-diffusion equation,
\begin{eqnarray}
    \frac{\partial \phi}{\partial t}=-u\frac{\partial \phi}{\partial x}-v\frac{\partial \phi}{\partial y}+\alpha\left(\frac{\partial^2 \phi}{\partial x^2}+\frac{\partial^2 \phi}{\partial y^2}\right),\ 0\leq x\leq 2\pi, -1\leq y \leq +1,\label{eq:2DAD}
\end{eqnarray}
where $u,v$ are the velocities in the $x$ and $y$ directions, respectively, in the case where the 
field is periodic
in the $x$ direction
and the Dirichlet condition is applied 
in the $y$ direction.
The Fourier series expansion in the $x$ direction and the Chebyshev series expansion in the $y$ direction for the $x$ and $y$ functions $\phi$ 
yield
\begin{eqnarray}
    \eqalign{\phi(t,x,y)=\sum_{n_x=0}^{N_x}\sum_{n_y=0}^{N_y}a_{n_xn_y}\cos{(n_xx)}T_{n_y}(y)\\
    +\sum_{n_x=1}^{N_x}\sum_{n_y=0}^{N_y}b_{n_xn_y}\sin{(n_xx)}T_{n_y}(y).}\label{eq:2Dexp}
\end{eqnarray}
Substituting
\Eref{eq:2Dexp} into \Eref{eq:2DAD} 
results in a minimization problem similar to that in \Sref{s:CSM}, which
can be handled by the Ising machine.

\section{Numerical experiments}\label{s:Ne}
\subsection{Error analysis}\label{s:EA}
\subsubsection{Analysis conditions}\label{s:Ac}

\begin{table}[!b]
    \caption{Analysis conditions in \Sref{s:EA}.}
    \begin{indented}
        \item[]\begin{tabular}{@{}ll}
            \br
            Velocity $u$ (and the Peclet number Pe) & $1,10$ \\
            Diffusivity $\alpha$ & $1$ \\
            Time step $\Delta t$ & $0.01$ \\
            Time range & $0 \leq t \leq 1$ \\
            Initial condition & $\phi\left(t=0\right)=1$ \\
            Boundary condition & $\phi\left(x=0\right)=0,\phi\left(x=1\right)=1$ \\
            Standard of boundary condition parameters $\gamma$ & 
            $\gamma=\mu\left(\left|A_{ij}\right|\right)$ \\
            Quantize scale $x_\mathrm{min},x_\mathrm{max}$ & $x_\mathrm{min}=-0.5,x_\mathrm{max}=1.5$ \\
            Ising machine & Fixstars Amplify AE (SA) \\
            Annealing time & $10000\ \mathrm{ms}$\\
            \br
        \end{tabular}
        \label{tab:ac}
    \end{indented}
\end{table}
\ 

\noindent
In \Sref{s:EA}, we perform error analysis through the computation of the steady state solution and initial-boundary value problems of the one-dimensional advection-diffusion equation.
The computational conditions in this section are shown in \Tref{tab:ac}.
The behavior of the solution of \Eref{eq:1DADeq} is determined by the Peclet number $\mathrm{Pe}=u/\alpha$, which is the ratio of the magnitude of the advection and diffusion effects, so the calculations are performed by varying only the flow velocity $u$, while the diffusivity is fixed at $\alpha=1$.
The parameter $\gl$ and $\gr$ that determines the weight of the boundary condition introduced in \Sref{s:LlsmQA} or \Sref{s:CSM} is based on the average of each element of the coefficient matrix obtained by \Eref{eq:Res}, 
\begin{eqnarray}
    \gamma&=\mu\left(\left|A_{ij}\right|\right) \nonumber \\ 
    &=\frac{1}{N_1N_2}\sum_{i=1}^{N_1}\sum_{j=1}^{N_2}\left|A_{ij}\right|,
\label{eq:gamma}
\end{eqnarray}
for the $N_1\times N_2$ matrix $\boldsymbol{A}$.
The scaling range
used in quantizing the continuous quantities 
is $[x_\mathrm{min}$, $x_\mathrm{max}]=[-0.5, 1.5]$.
Fixstars Amplify AE (Fixstars Amplify Corporation 2024) is used as the simulated annealing machine.
The annealing time is the actual time used for a single QUBO calculation. 
In both quantum and simulated annealing, the longer the time spent per computation, the more searches can be performed, which may improve the accuracy of QUBO.
However, the longer the annealing time, the less advantage it has over classical computation schemes, so we here set a relatively short annealing time.

\subsubsection{Comparison 
between FDM-based and CSM-based simulated annealings}\label{s:CwFDM}
\ 

\begin{figure}
    \centering
    \includegraphics[width=\linewidth]{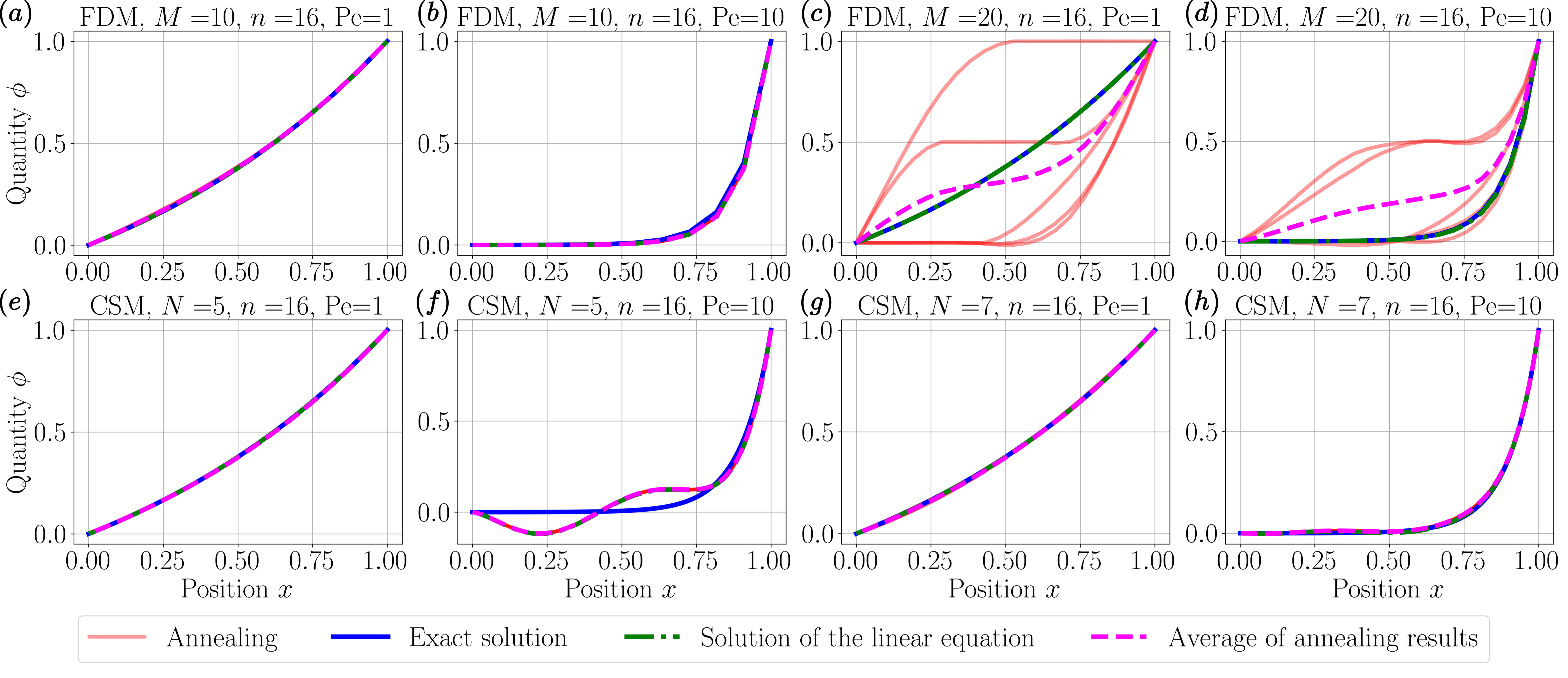}
    \caption{Steady state solutions by the simulated annealing with  the FDM and the CSM.
    $(a)$ FDM, $M=10$, $\mathrm{Pe}=1$.
    $(b)$ FDM, $M=10$, $\mathrm{Pe}=10$.
    $(c)$ FDM, $M=20$, $\mathrm{Pe}=1$.
    $(d)$ FDM, $M=20$, $\mathrm{Pe}=10$.
    $(e)$ CSM, $N=5$, $\mathrm{Pe}=1$.
    $(f)$ CSM, $N=5$, $\mathrm{Pe}=10$.
    $(g)$ CSM, $N=7$, $\mathrm{Pe}=1$.
    $(h)$ CSM, $N=7$, $\mathrm{Pe}=10$.
        The parameter for quantization is $n=16$ for all cases.
    Exact solutions (blue lines) and solutions of the linear equations (green lines) almost overlap in every 
subfigure
    except $(f)$.
    }
    \label{fig:FDMvCSMs}
\end{figure}

\noindent
The 
steady state 
solutions obtained by
the FDM-based and 
CSM-based simulated annealing are shown in \Fref{fig:FDMvCSMs}.
Here, the simulated annealing is a 
stochastic
computation scheme, so the steady state is computed five times each
using different random seeds.
In each 
subfigure, the red line denotes the result of simulated annealing, the blue line denotes the analytical solution, i.e.,
\begin{equation}
    \phi(x)=\frac{\exp{\left(\mathrm{Pe}\,x\right)-1}}{\exp{\left(\mathrm{Pe}\right)-1}},
\end{equation}
the green line denotes the 
classical numerical
solution of 
\Eref{eq:LE}, and the magenta line denotes the average of the simulated annealing results.
\Fref{fig:FDMvCSMs} $(a)$--$(d)$ show the 
FDM-based results,
 where $M$ denotes the number of calculation points.
\Fref{fig:FDMvCSMs} $(e)$--$(h)$ are the CSM-based results,
 where $N$ denotes
 the expansion order.
In this section, the Chebyshev--tau method is 
used for the treatment of boundary conditions in CSM.
For all 
cases,
the precision of quantization of continuous quantities 
(see, \Eref{eq:exp_of_x})
is set to $n=16$.

\Fref{fig:FDMvCSMs} shows that  the FDM-based annealing computes accurately for both Pe $=1$ and $10$ when $M=10$.
On the other hand, when $M=20$, 
each trial of
FDM-based annealing
significantly deviates from
the analytical solution, and the averaged
profile also deviates significantly from the analytical 
one.
This indicates that the accuracy of the FDM-based annealing worsens as the number of calculation points increases, and the variance of the obtained solution increases.
This is likely 
attributed to
the increase in the number of variables, which increases the scale of the optimization problem
and makes
it more difficult to find a solution.
The number of bits used in simulated annealing is given by $Mn$, since each of the $M$ continuous variables is quantized with $n$ bits.
Therefore, the problem size of the FDM-based computation in \Fref{fig:FDMvCSMs} is $Mn=160$ for $(a)$ and $(b)$, and $Mn=320$ for $(c)$ and $(d)$.
In $(c)$ and $(d)$, the problem size increases, and the size of the search area increases by a factor of $2^{160}$.
Therefore, the optimal solution in $(c)$ and $(d)$ cannot be reached in $10000\ \mathrm{ms}$.

\Fref{fig:FDMvCSMs} $(e)$--$(h)$ 
show the results obtained by the methods based on
the CSM.
The simulated annealing results 
are found to be in good agreement with the conventional solution of the linear equation $\boldsymbol{Ax}=\boldsymbol{b}$, which also suggests that
the deviation from the exact solution at $N=5$ is simply due to the 
insufficient order of expansion.
Even when $N$ is large, 
accurate 
solutions are obtained unlike
the FDM-based one.
Since the spectrum $a_i$ exists from $0$th to $N$th order, 
each approximated using
$n$ cubits, the problem size of QUBO is $(N+1)n$.
The problem size of the CSM-based computation
is $(N+1)n=96$ for
the cases presented in \Fref{fig:FDMvCSMs}
 $(e)$ and $(f)$, and $(N+1)n=112$ for the cases presented in
 \Fref{fig:FDMvCSMs}
 $(g)$ and $(h)$, which 
 are substantially
 smaller than 
 those
 of the corresponding FDM-based computations.
Therefore, the search is simpler, and more accurate solutions are output compared to those computed with the FDM-based annealing.

\begin{figure}[t]
    \centering
    \includegraphics[width=\linewidth]{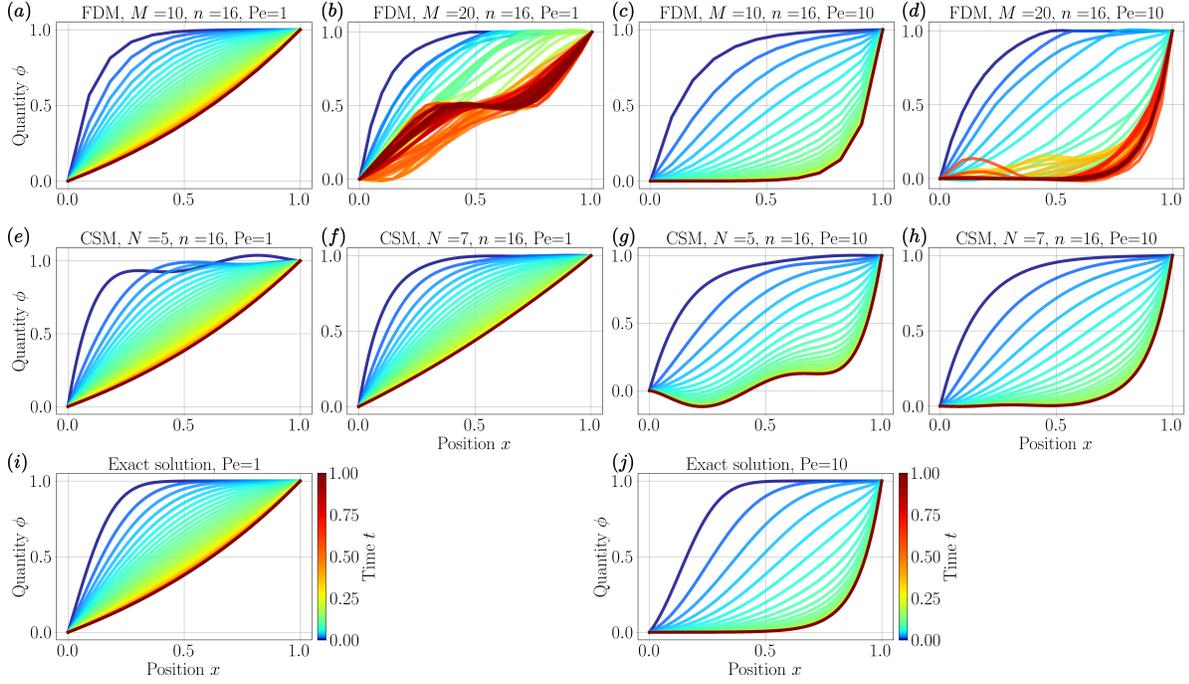}
    \caption{
    Solutions for the unsteady problems.
    $(a)$ FDM, $M=10$, $\mathrm{Pe}=1$.
    $(b)$ FDM, $M=20$, $\mathrm{Pe}=1$.
    $(c)$ FDM, $M=10$, $\mathrm{Pe}=10$.
    $(d)$ FDM, $M=20$, $\mathrm{Pe}=10$.
    $(e)$ CSM, $N=5$, $\mathrm{Pe}=1$.
    $(f)$ CSM, $N=7$, $\mathrm{Pe}=1$.
    $(g)$ CSM, $N=5$, $\mathrm{Pe}=10$.
    $(h)$ CSM, $N=7$, $\mathrm{Pe}=10$.
    $(i)$ Exact solution, $\mathrm{Pe}=1$.
    $(j)$ Exact solution, $\mathrm{Pe}=10$.
    The parameter for quantization is $n=16$ for all cases.
    }
    \label{fig:FDMvCSMus}
\end{figure}

Next, let us 
compare the FDM-based and  
CSM-based simulated annealings through the computation of the 
unsteady
problems.
For these cases, each computation 
is performed once.
\Fref{fig:FDMvCSMus} shows the results 
obtained
using the FDM-based and 
 CSM-based annealings
 for different Peclet numbers Pe
 using different numbers of computational points $M$ (for FDM) or the expansion order $N$ (for CSM).
The analytical solution is given as
\begin{equation}
    \eqalign{\phi(t,x)=\frac{\exp{\left(\mathrm{Pe}\,x\right)-1}}{\exp{\left(\mathrm{Pe}\right)-1}}\cr
    +\exp{\left[\frac{\mathrm{Pe}}{2}\left(x-\frac{u}{2}t\right)\right]}\sum_{k=0}^{\infty}\frac{8n\pi}{\mathrm{Pe}^2+4n^2\pi^2}\exp{\left(-\alpha n^2 \pi^2 t\right)}\sin{\left(n\pi x\right)},}\label{eq:ASofUS}
\end{equation}
and we use sums up to the 100th order.
In all calculations in \Fref{fig:FDMvCSMus}, the precision of quantization of continuous quantities is set to $n=16$, as in the 
case of the
steady 
problem.

\Fref{fig:FDMvCSMus} shows that the 
FDM-based annealing significantly impairs the accuracy
 as in the steady state.
In particular, in the case of $M=20$, which could not be calculated with high accuracy even 
for
the steady 
problem, the error looks to accumulate also in its time integration.
On the other hand, \Fref{fig:FDMvCSMus} $(e)$--$(h)$ calculated with the CSM-based annealing
show
reasonable
agreement with the exact solution, although the 
error increases when the Peclet number $\mathrm{Pe}$ is large.
Although the accumulation of error due to the time integration is present in this case, too, its amount looks to be relatively small because the error in each time step is sufficiently small.

\subsubsection{Effect of computational parameters $N$ and $n$}\label{s:Eop}

\noindent
In this section, we investigate the effect of the expansion order $N$ of the Chebyshev spectral method and the precision of quantization $n$ of the spectrum on the computation with the Ising machine.
In classical 
computations,
the larger $N$ or $n$
will result in
 the higher the accuracy of the calculation.
However, 
as we have seen in the FDM-based annealing,
the increase in the size of the calculation in the Chebyshev spectral method will also 
deteriorate the accuracy.
In this subsection, we examine the impact of these parameters on the 
CSM-based annealing in more detail.

\begin{figure}[t]
    \centering
    \includegraphics[width=\linewidth]{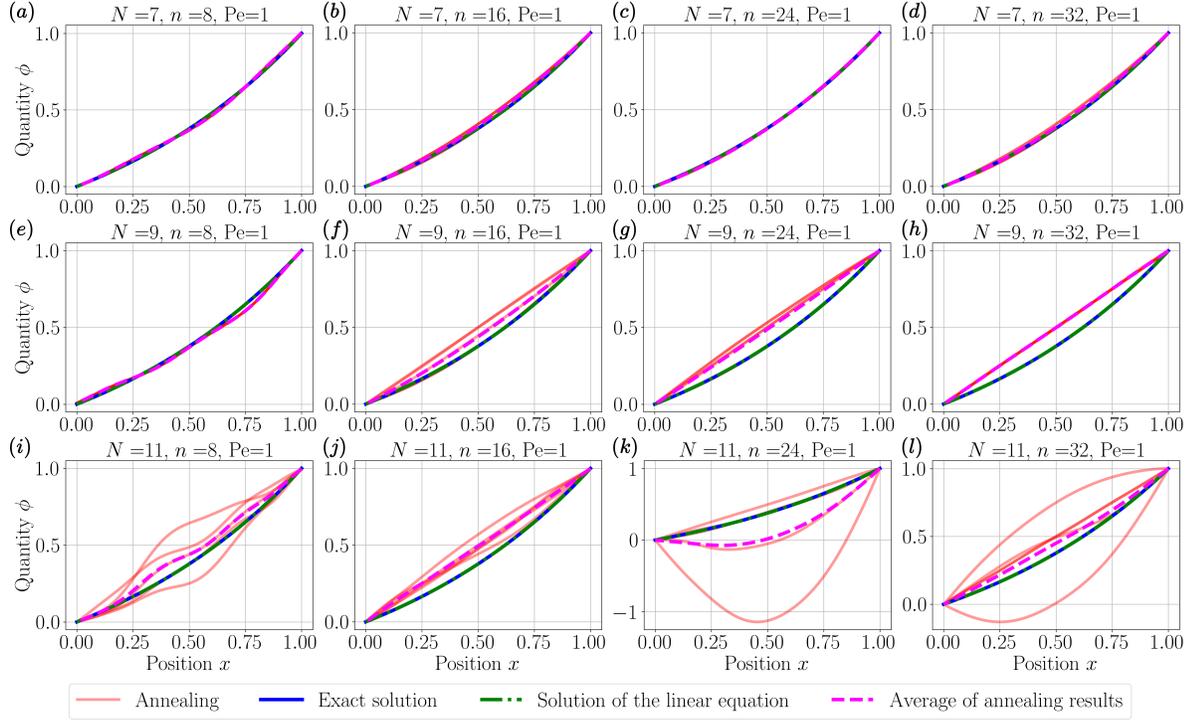}
    \caption{Steady state solution by simulated annealing at $\mathrm{Pe}=1$, calculated using the Chebyshev-tau method.
    Row, different expansion orders:
    $(a)$--$(d)$, $N=7$;
    $(e)$--$(h)$, $N=9$;
    $(i)$--$(l)$, $N=11$.
    Column, different numbers of quantization:
    $(a) (e) (i)$, $n=8$;
        $(b) (f) (j)$, $n=14$;
            $(c) (g) (k)$, $n=24$;
                $(d) (h) (l)$, $n=32$;
    }
    \label{fig:Nn1s}
\end{figure}

\Fref{fig:Nn1s} 
shows the 
CSM-based results 
for the
steady 
problem at Pe $=1$ obtained using
different $N$ and $n$.
For $N=7$, the annealing results
are in good agreement with
 the analytical solution (blue line) or the solution of the linear equation (green line), indicating that the computation is accurate.
However, in the case of $N=9$, some annealing solutions deviate from the correct solutions.
Also, in the case of $N=11$, the output is almost never accurate, and averaging does not improve the results.
This suggests that the increase in computational scale also deteriorates the results of 
the CSM-based annealing.
In contrast,
\Fref{fig:Nn1s} 
shows that changing $n$ does not 
always
deteriorate the solution.
Note that a similar dependence on $N$ and $n$ has been confirmed for the case of Pe $= 10$, although not shown here for brevity.

\begin{figure}[t]
    \centering
    \includegraphics[width=0.8\linewidth]{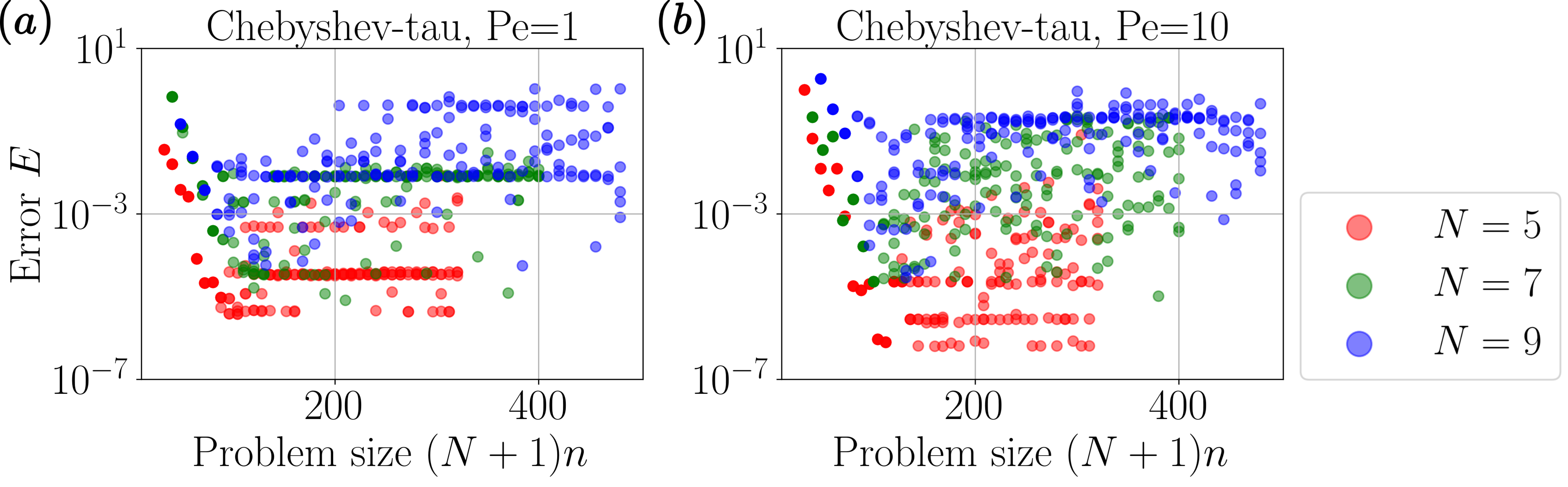}
    \caption{Relations between the problem sizes $(N+1)n$ and the computational error $E$.
    $(a)$ 
    $\mathrm{Pe}=1$;
    $(b)$ 
    $\mathrm{Pe}=10$.}
    \label{fig:scatCSMNn}
\end{figure}

Since the computational scale is expressed as $(N+1)n$, the computational scale should increase as well when $n$ is increased.
Therefore, differences in the effects of $N$ and $n$ on calculation accuracy need to be investigated.
\Fref{fig:scatCSMNn} shows the relationship between the scale of computation and computational accuracy.
\Fref{fig:scatCSMNn} $(a)$ and $(b)$ show the results of calculations for Pe $=1$ and Pe $=10$, respectively, for $N=7, 9$, and $11$, while varying $n$ in the range of $4\leq n \leq 40$.
For each calculation condition, 
five calculations were performed, and the results of each calculation are plotted one by one.
The horizontal axis is the scale of computation $(N+1)n$, and the vertical axis is the mean squared error $E$ which is computed as,
\begin{eqnarray}
    E=\sum_{k=0}^{K}\left[\phi_{\mathrm{Annealing}}(\xi_k)-\phi_{\mathrm{Linear}}(\xi_k)\right]^2w_k,
\end{eqnarray}
where $\phi_{\mathrm{Annealing}}(\xi_k)$ is the value of $\phi$ at the $k$-th Chebyshev 
collocation
point $\xi_k=\cos{\left(k\pi/
N
\right)}$ in the annealing solution, $\phi_{\mathrm{Linear}}(\xi_k)$ is the value of $\phi$ at $\xi_k$ in the linear equation solution, and $w_k$ is the weight function in the Crenshaw--Curtis quadrature~\cite{CC1960,S2013}.
\Fref{fig:scatCSMNn} shows that the accuracy 
varies greatly with the value of 
$N$.
On the other hand, the accuracy does not change significantly when $n$ is changed.

This deterioration 
in
accuracy 
for larger $N$
is likely 
related to
the condition number of the 
matrix $\boldsymbol{A}$,
which is
defined by
\begin{eqnarray}
    \kappa(\boldsymbol{A})=\left\|\boldsymbol{A}\right\|\left\|\boldsymbol{A}^{-1}\right\|,\label{eq:def_o_cond}
\end{eqnarray}
and is an indicator of the difficulty of solving linear equations~\cite{P1956}.
Here, $\left\|\boldsymbol{A}\right\|$ is the operator norm of matrix $\boldsymbol{A}$.
Due to submultiplicity of the operator norms, $\left\|\boldsymbol{AB}\right\|\leq\left\|\boldsymbol{A}\right\|\left\|\boldsymbol{B}\right\|$,
\begin{eqnarray}
    \boldsymbol{A}\left(\boldsymbol{x}+\delta\boldsymbol{x}\right)-\boldsymbol{b}=\delta\boldsymbol{r},\\
    \frac{\left\|\delta\boldsymbol{\boldsymbol{x}}\right\|}{\left\|\boldsymbol{x}\right\|}\leq\left\|\boldsymbol{A}\right\|\left\|\boldsymbol{A}^{-1}\right\|\frac{\left\|\delta\boldsymbol{\boldsymbol{r}}\right\|}{\left\|\boldsymbol{b}\right\|}=\kappa(\boldsymbol{A})\frac{\left\|\delta\boldsymbol{\boldsymbol{r}}\right\|}{\left\|\boldsymbol{b}\right\|},
\end{eqnarray}
which suggests that the error $\left\|\delta\boldsymbol{x}\right\|$ is bounded by a constant multiple of the residual $\left\|\delta\boldsymbol{r}\right\|$.
Thus, when the coefficient matrix $\boldsymbol{A}$ has a larger condition number $\kappa(\boldsymbol{A})$, the computational error $\left\|\delta\boldsymbol{\boldsymbol{x}}\right\|/\left\|\boldsymbol{x}\right\|$ is larger 
for a given normalized residual $\left\|\delta\boldsymbol{\boldsymbol{r}}\right\|/\left\|\boldsymbol{b}\right\|$.
\Tref{tab:kappaNn} shows the condition number of the coefficient matrix $\boldsymbol{A}$ for each $N$.
Since the condition number increases as $N$ increases, solving linear equations becomes more difficult, and the accuracy of the calculation decreases.
On the other hand, since $n$ does not affect the change in the condition numbers, it has the effect of increasing the computational scale of combinatorial optimization, but its effect is smaller than the increase in the condition numbers and does not degrade the accuracy of the computation.


\begin{table}[t]
    \caption{Condition numbers $\kappa(\boldsymbol{A})$ 
    for different expansion order $N$ 
    in the 
    computations of steady state solution at $\mathrm{Pe}=1$.
    The boundary weight $\gamma$ is defined in \Eref{eq:gamma}.}
    \begin{indented}
        \item[]\begin{tabular}{@{}llll}
            \br
            Method & $\gl,\gr$ & $N$ & $\kappa(\boldsymbol{A})$ \\
            \mr
            Chebyshev--tau & $\gamma$ & 5 & 61.0 \\
             & & 7 & 203.4 \\
             & & 9 & 508.6 \\
             & & 11 & 1068.3 \\
            \br
        \end{tabular}
        \label{tab:kappaNn}
    \end{indented}
\end{table}

\subsubsection{Effect of the weights of boundary conditions $\gl$ and $\gr$}\label{s:gamma}
\ 

\begin{figure}[b]
    \centering
    \includegraphics[width=\linewidth]{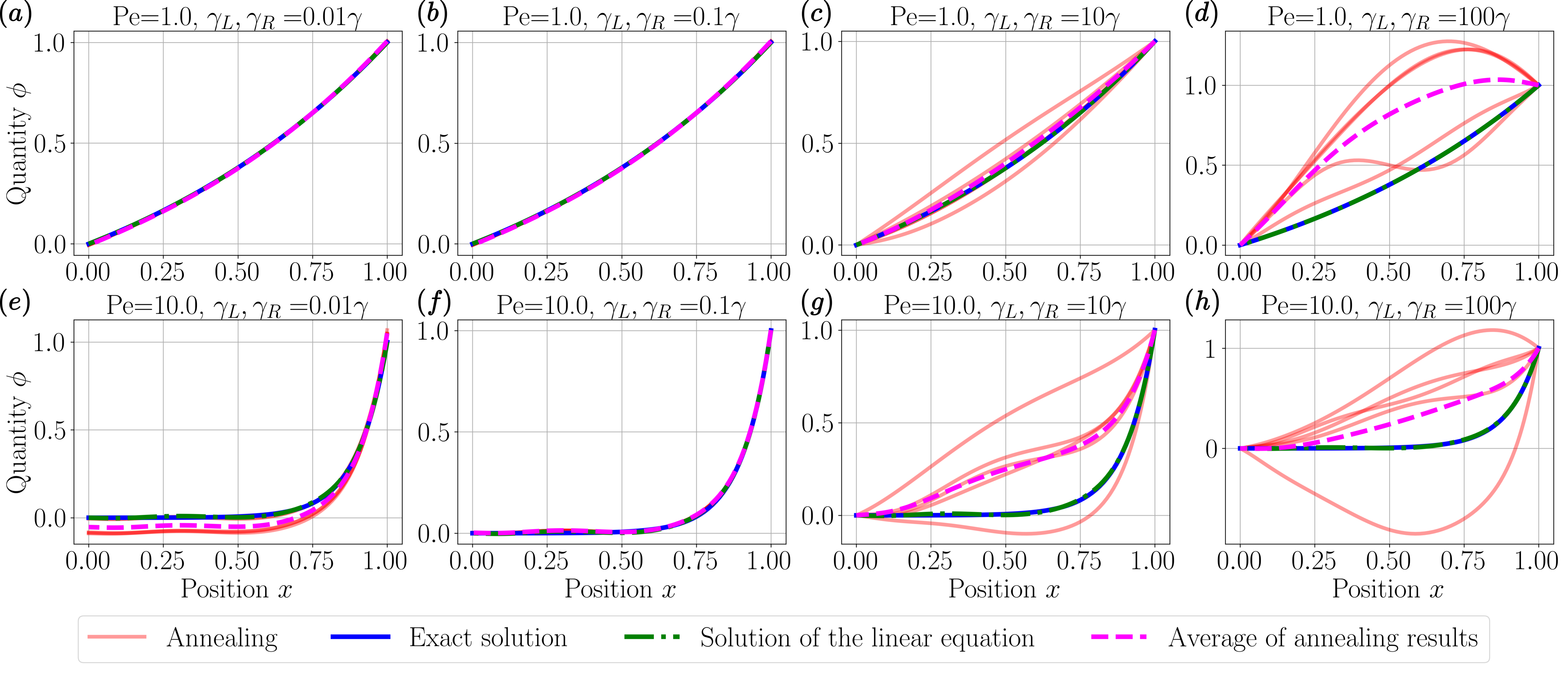}
    \caption{Steady state solution obtained by Chebyshev-tau-based simulated annealing 
    with different boundary weights $\gl$ and $\gr$.
    The reference boundary weight $\gamma$ is defined in \Eref{eq:gamma}.
$(a)$--$(d)$, $\mathrm{Pe}=1$;
$(e)$--$(h)$, $\mathrm{Pe}=10$.
$(a) (e)$, $\gl,\gr=0.01\gamma$;
$(b) (f)$, $\gl,\gr=0.1\gamma$;
$(c) (g)$, $\gl,\gr=10\gamma$;
$(d) (g)$, $\gl,\gr=100\gamma$.
    The expansion and quantization parameters are $N=7$ and $n=16$. 
    }
    \label{fig:SSgamma1}
\end{figure}

\noindent
In this section, we examine the effect of the boundary condition weight $\gl$ and $\gr$.
\Fref{fig:SSgamma1} shows 
the steady state solution 
obtained by the Chebyshev-tau-based simulated annealing with
different $\gl$ and $\gr$.
When $\gl=\gr=0.01\gamma$,
 the boundary conditions have little effect on the objective function in QUBO form
so that
the solutions 
that do not 
satisfy
the boundary conditions are obtained.
In contrast, when 
$\gl=\gr=101\gamma$ and $100\gamma$,
the influence of the boundary conditions is 
obviously too large.

There is no analytical guideline for the optimal value of $\gl$ and $\gr$, which should neither be too large nor excessively small.
However,
as in \Sref{s:Eop}, the condition number of the coefficient matrix can be used as the basis to indicate 
a
suitable value.
\Tref{tab:kappagamma} shows the condition numbers for steady state calculations for different values of $\gl,\gr$.
We can see that the condition number is larger for small and large values of $\gl$ and $\gr$.
When $\gl$ and $\gr$ are close to $\gamma$ defined in \Eref{eq:gamma}, the condition number tends to be smaller.

\begin{table}[t]
    \caption{Condition numbers $\kappa(\boldsymbol{A})$ for 
    different values of $\gl$ and $\gr$
    in the 
    steady state
    computation 
    with the Chebyshev-tau method. $N=7$, $\mathrm{Pe}=1$.}
    \begin{indented}
        \item[]\begin{tabular}{@{}ll}
            \br
            $\gl,\gr$ & $\kappa(\boldsymbol{A})$ \\
            \mr
            $0.01\gamma$ & 1123.4 \\
            $0.1\gamma$ & 216.7 \\
            $\gamma$ & 203.4 \\
            $10\gamma$ & 408.0 \\
            $100\gamma$ & 3836.7 \\
            \br
        \end{tabular}
        \label{tab:kappagamma}
    \end{indented}
\end{table}


\subsubsection{Comparison 
among different
formulations in Chebyshev spectral method}\label{s:CofmCSM}
\ 

\noindent
In this section, we investigate the differences 
among different
formulations of the Chebyshev spectral methods, i.e., tau, Galerkin, and collocation methods detailed in Appendix A, and the penalty method introduced by \Eref{eq:fCSM}.
Although we have examined two Peclet numbers, Pe $= 1$ and 10, here we show the results of Pe $=1$ cases only, because the trends are essentially the same.

\begin{figure}[!t]
    \centering
    \includegraphics[width=\linewidth]{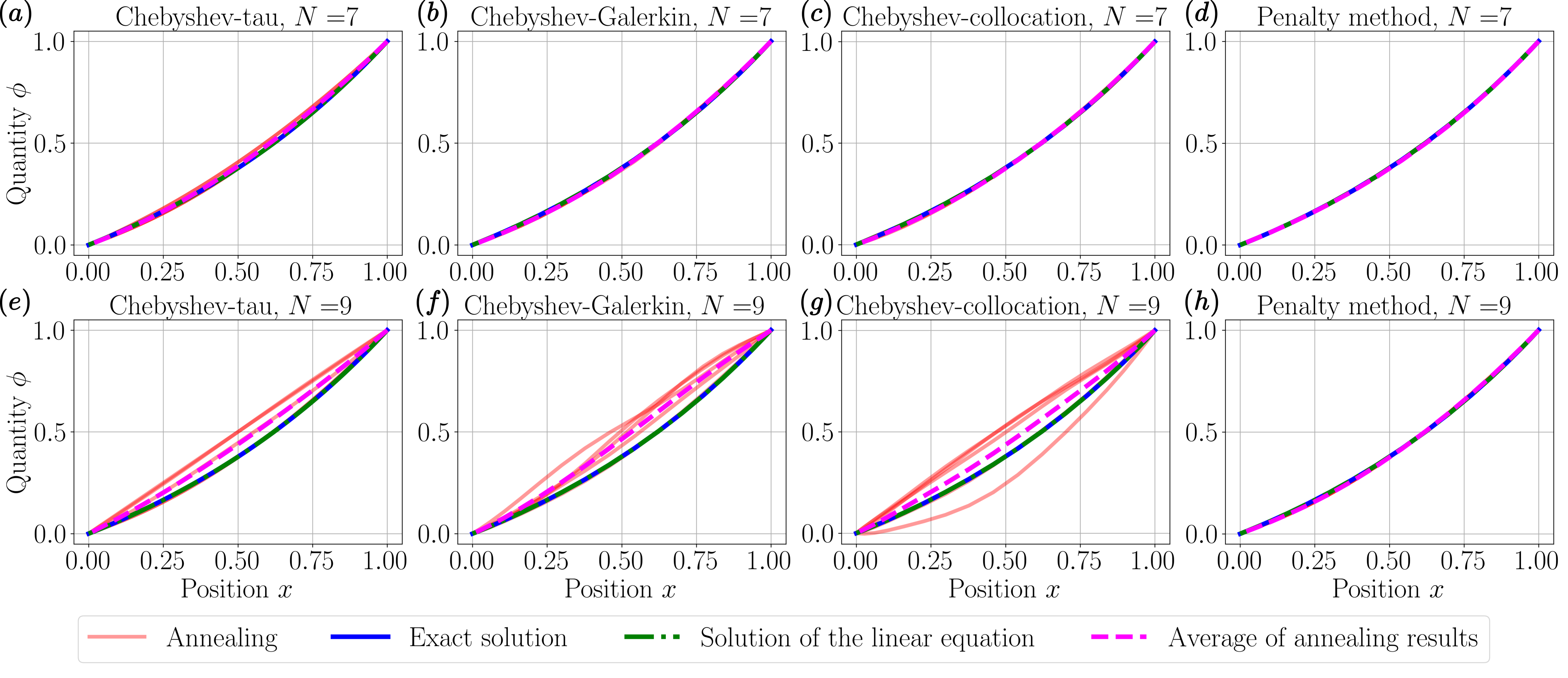}
    \caption{Steady state solution 
    obtained by simulated annealing with different formulations of Chebyshev methods.
    $\mathrm{Pe}=1$, $n=16$.
	$(a)$--$(d)$, $N=7$;
	$(e)$--$(h)$, $N=9$.
	$(a) (d)$, tau method;
	$(b) (e)$, Galerkin method;
    $(c) (f)$, collocation method;
	$(d) (g)$, penalty method.
	}
    \label{fig:formSS1}
\end{figure}

\Fref{fig:formSS1} shows the profiles 
obtained using the different methods.
These results show that the 
accuracy is not 
deteriorated by the penalty method
when $N$ is increased, 
while the conventional three methods fail to produce reasonable profiles at $N=9$.
This results suggest that the penalty
method 
is more suitable for the computations on Ising machines than the conventional three methods.
Its superiority is also indicated by the correlation between the objective function of annealing and the computational error, i.e.,
Spearman's rank correlation coefficients $\rho$ between the objective function and the computational error for the tau method and the alternative method, 
shown in \Tref{tab:rhoTvA}.
From \Tref{tab:rhoTvA}, we can see that the 
penalty method has a closer relationship between the computational error and the monotonically increasing function of the objective function.
When solving by annealing, it is desirable to have a monotonically increasing relationship between the computational error to be minimized and the objective function that is actually minimized.
These results indicate that the alternative method is the good computation scheme using annealing.

\begin{table}[t]
    \caption{Spearman's rank correlation coefficients $\rho$ between the objective function $f$ and the computational error $E$ in the 
    steady 
    solutions at
    $\mathrm{Pe}=1$.}
    \begin{indented}
        \item[]\begin{tabular}{@{}lll}
            \br
            Formulation method & $N$ & $\rho$ \\
            \mr
            Tau method & 7 & 0.696 \\
             & 9 & 0.593 \\
             & 11 & 0.550 \\
            Penalty method & 7 & 0.924 \\
             & 9 & 0.779 \\
             & 11 & 0.677 \\
            \br
        \end{tabular}
        \label{tab:rhoTvA}
    \end{indented}
\end{table}

\subsection{Extension to two-dimensional problems}\label{s:Et2D}

In this section, we consider solving the two-dimensional advection-diffusion equation with an Ising machine.
As shown in \Sref{s:FSM}, the calculations are performed using both  the CSM and the FSM.

The computational conditions are shown in \Tref{tab:ac2D}, where $u$ is the flow velocity in the $x$-direction, $v$ is the flow velocity in the $y$-direction, $N_x$ is the expansion order of the FSM in the $x$-axis direction, and $N_y$ is the expansion order of the CSM in the $y$-axis direction.
There are $4N_x+2$ first-order equations for the boundary conditions, for each of which the boundary condition parameter $\gamma$ can be defined.
For simplicity, the value of $\gamma$ is the average of the absolute values of the coefficients in the coefficient matrix $\boldsymbol{A}$ of the first-order equation for the residuals, as is the case with $\gamma$ for one-dimensional problems.

\begin{table}[]
    \caption{Analysis conditions in \Sref{s:Et2D}.}
    \begin{indented}
        \item[]\begin{tabular}{@{}ll}
            \br
            Velocity $u$ ($=$ Peclet number Pe) & $1$ \\
            Diffusivity $\alpha$ & $1$ \\
            Time step $\Delta t$ & $0.01$ \\
            Initial condition & $\phi\left(t=0\right)=1$ \\
            Boundary condition & $\phi\left(y=-1\right)=0.5\left(1-\sin x\right)$, \\
             & $\phi\left(y=+1\right)=0$ \\
            Boundary condition parameters $\gamma$ & 
            $\gamma=\mu\left(\left|A_{ij}\right|\right)$ \\
            Quantize scale $x_\mathrm{min},x_\mathrm{max}$ & $x_\mathrm{min}=-0.5,x_\mathrm{max}=1.5$ \\
            Ising machine & Fixstars Amplify AE (SA) \\
            Annealing time & $10000\ \mathrm{ms}$\\
            \br
        \end{tabular}
        \label{tab:ac2D}
    \end{indented}
\end{table}

\begin{figure}
    \centering
    \includegraphics[width=0.8\linewidth]{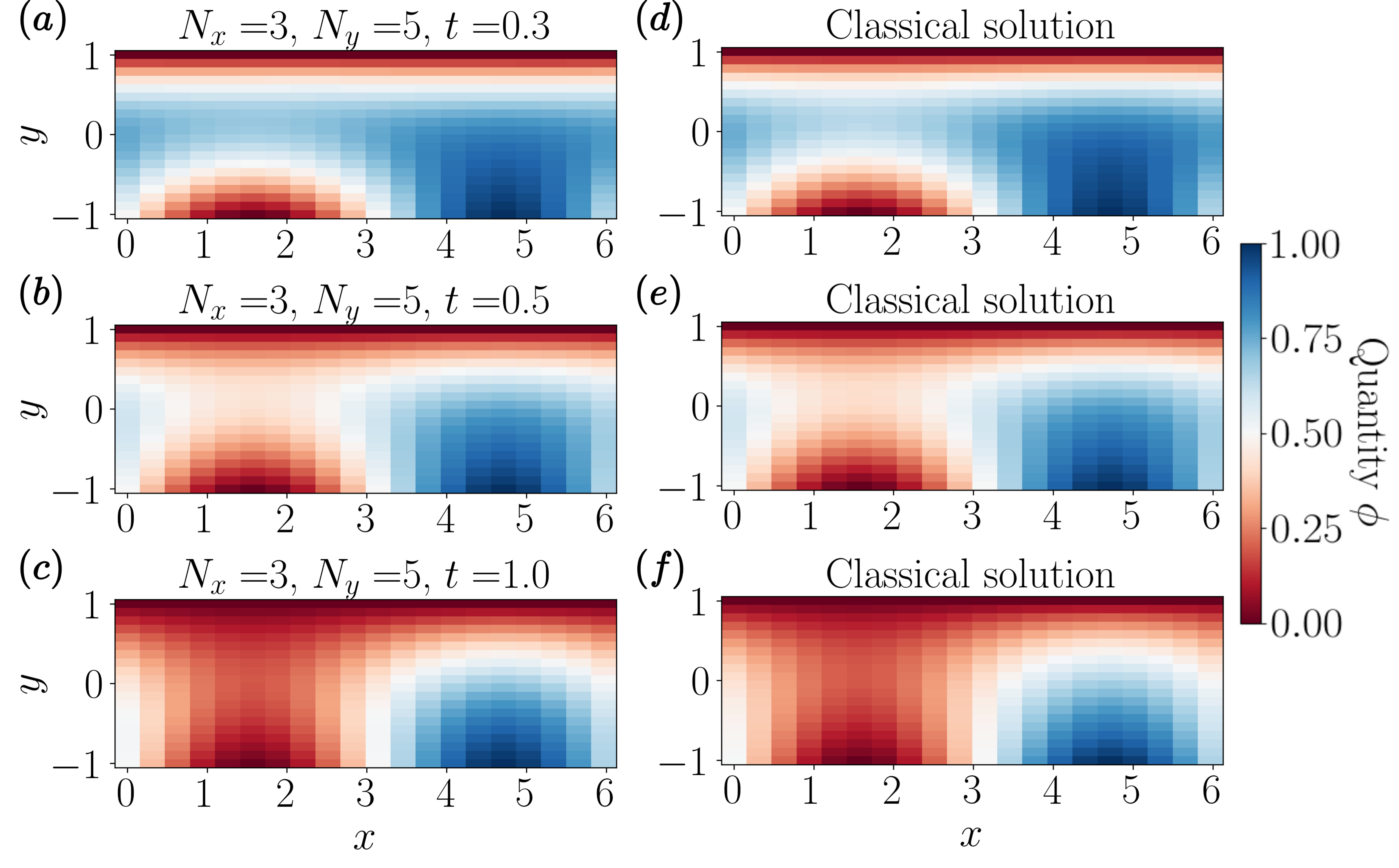}
    \caption{Unsteady 
    solutions of the two dimensional problem 
    obtained by Chebyshev-tau-based simulated annealing with $(N_x,N_y,n)=(3,5,16)$.
$(a)$--$(c)$ Simulated annealing;
$(d)$--$(f)$ Conventional linear solution.
$(a) (d)$, $t=0.3$;
$(b) (e)$, $t=0.5$;
$(c) (f)$, $t=1.0$.
    }
    \label{fig:2D35}
\end{figure}

The results of unsteady 
computations with
the expansion orders $(N_x,N_y)=(3,5)$ are shown in \Fref{fig:2D35}
and compared with the results obtained by the conventional linear solution.
From 
\Fref{fig:2D35}
the results are satisfactory, confirming
that 
the extension to two-dimensional problem is straightforward.
For 
this
problem, the Fourier 
series
have $N_x+1$ cosine series and $N_x$ sine series, for a total of $2N_x+1$, and the Chebyshev spectrum has $N_y+1$, for a total of $(2N_x+1)(N_y+1)$ variables.
Therefore, the problem size of QUBO is $(2N_x+1)(N_y+1)n=672$ in \Fref{fig:2D35}.
This
suggests
 that the problem can be computed accurately despite the scale of the optimization problem is much larger than that of the one-dimensional problem.
 The reason for this can be explained again by the
condition number
 shown in \tref{tab:kappa2D}.
The condition numbers for this problem
are smaller than 
those
in \Sref{s:EA}, indicating 
relatively small
errors.

\begin{table}[t]
    \caption{Condition numbers $\kappa(\boldsymbol{A})$ for each $N$ in the 
    two-dimensional problem with $u=1, v=1, \alpha=1, \Delta t=0.01$.}
    \begin{indented}
        \item[]\begin{tabular}{@{}llll}
            \br
            Formulation method & $\gl,\gr$ & $(N_x,N_y)$ & $\kappa(\boldsymbol{A})$ \\
            \mr
            Chebyshev--tau & $\gamma$ & $(3,5)$ & 3.30 \\
            \br
        \end{tabular}
        \label{tab:kappa2D}
    \end{indented}
\end{table}

\subsection{Implementation on quantum annealers}\label{s:IiQA}

In \Sref{s:EA} and \Sref{s:Et2D}, we performed numerical experiments using simulated annealing.
In this section, we perform steady state 
computations
using the Advantage quantum system (Advantage), the world's first commercial quantum annealer developed by D-Wave Systems (D-Wave).

\Tref{tab:acQA} shows the 
computational
conditions.
What is specific about a quantum annealer is the
number of read on the bottom of table, which specifies
the number of calculations
per an output, at which interval the quantum annealer
returns the solution with the lowest energy among multiple calculations.
In the present study, this
number of read is set to $500$.

\begin{table}[b!]
    \caption{
    Computational
    conditions in \Sref{s:IiQA}.}
    \begin{indented}
        \item[]\begin{tabular}{@{}ll}
            \br
            Velocity $u$ & $1$ \\
            Diffusivity $\alpha$ & $1$ \\
            Boundary condition & $\phi\left(x=0\right)=0,\phi\left(x=1\right)=1$ \\
            Boundary condition parameters $\gl,\gr$ & 
            $\gl,\gr=\mu\left(\left|A_{ij}\right|\right)$ \\
            Quantize scale $x_\mathrm{min},x_\mathrm{max}$ & $x_\mathrm{min}=-0.5,x_\mathrm{max}=1.5$ \\
            Ising machine & Advantage 4.1 (QA) \\
            Number of read & $500$\\
            \br
        \end{tabular}
        \label{tab:acQA}
    \end{indented}
\end{table}

\begin{figure}
    \centering
    \includegraphics[width=\linewidth]{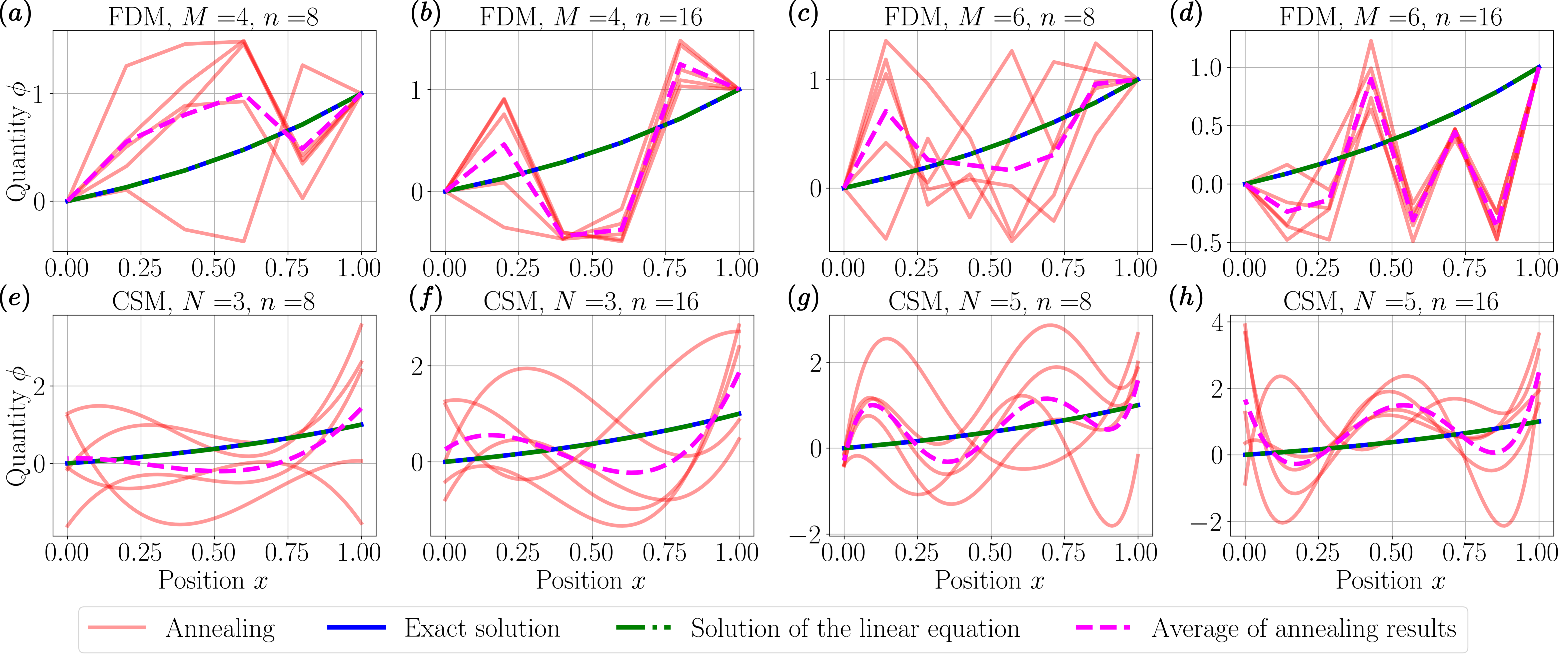}
    \caption{Steady state solutions obtained by the quantum annealer (D-Wave Systems) with the FDM and 
    CSM-based methods.
    $(a)$ FDM, $M=4$, $n=8$.
    $(b)$ FDM, $M=4$, $n=16$.
    $(c)$ FDM, $M=6$, $n=8$.
    $(d)$ FDM, $M=6$, $n=16$.
    $(e)$ CSM, $N=3$, $n=8$.
    $(f)$ CSM, $N=3$, $n=16$.
    $(g)$ CSM, $N=5$, $n=8$.
    $(h)$ CSM, $N=5$, $n=16$.
    }
    \label{fig:QA}
\end{figure}

The results 
obtained on
the Advantage quantum annealer are shown in \Fref{fig:QA}.
Unfortunately, both the FDM-based method and the CSM-based method resulted
in very inaccurate results
even though the problem size is 
small compared to 
the problem considered in \Sref{s:EA}.
The cause of this inaccuracy 
is likely
due to 
the
structural differences between the actual quantum annealer and 
its simulator,
in addition to the effects of noise during the calculation and the uncertainties in the observation of the quantum system.
When the linear least squares method is implemented on an Ising machine, if the coefficient matrix $\boldsymbol{A}$ is a dense matrix, the matrix of interactions $\boldsymbol{J}$ in the Ising model will be a dense matrix as well.
In this case, the Ising model has a structure called an all-pairwise coupling graph, in which every qubit used in the computation has an interaction with every other qubit.
The number of interactions increases by $\Or(\nu^2)$ with the computational size $\nu$ of the Ising model.
Therefore, developing the quantum annealer that can implement all of them becomes more difficult with larger computers.
For that reason, the Advantage quantum annealer employs a sparse connectivity graph in which each qubit interacts only with a limited number of nearby qubits.
In particular, Advantage 4.1
we used in the present study adopts
a structure called a Pegasus connectivity graph, in which each qubit has 15 interactions.
Such a sparse connectivity graph simplifies physical implementation because the number of interactions is only $\Or(\nu)$.
On the other hand, when more interactions have to be considered than the sparse connectivity graph can handle, some alternative bits are needed.
Alternative bits are introduced to connect distant qubits that cannot be connected.
The method of embedding an all-pairwise coupling graph into the sparse connectivity graph by placing alternative bits is called Minor-Embedding (ME).
Since the number of alternative bits increases by $\Or(\nu^2)$, the size of the Ising model with the quantum annealer is considerably larger than the Ising model that can be created by transforming the problem to be solved (Shirai {\it et al} 2020).

\begin{table}[b!]
    \caption{The number of alternative bits per logic bit.}
    \begin{indented}
        \item[]\begin{tabular}{@{}ll}
            \br
            Number of logic bits & Number of alternative bits per logic bit \\
            \mr
            32 & 3.33 \\
            48 & 5.19 \\
            64 & 6.74 \\
            96 & 11.1 \\
            \br
        \end{tabular}
        \label{tab:altbit}
    \end{indented}
\end{table}

\Tref{tab:altbit} shows the number of alternative bits per logic bit when embedding the all-pairwise coupling graph in the Pegasus connectivity graph.
\Tref{tab:altbit} is the average of five calculations using the ME algorithm provided by D-Wave.
Since the ME algorithm uses a heuristic method, there is no need to have these numbers of alternative bits, but the size of the physical graph is larger than that of the logical graph.
Therefore, the output is far from the exact solution, even though the problem size is excessively small compared to the computation using simulated annealing.

\section{Summary and outlook}\label{s:Conc}

In order to clarify the current possibilities and limitations of fluid flow simulations on quantum annealers, we performed several numerical experiments
by taking simple example problems.
In particular, spectral methods with the linear least squares method were implemented on the Ising machines using the simulated annealing and the quantum annealing, and steady state and unsteady solutions of one-dimensional advection-diffusion equation were calculated to investigate efficiency of each calculation method.
Validation using simulated annealing showed that the annealing results with the CSM were more accurate than those with the FDM due to the smaller number of required variables.

The effects of calculation parameters such as the expansion order $N$, precision $n$, and weights of boundary conditions $\gl,\gr$ were also investigated.
The results showed that, unlike classical computations, increasing $N$ and $n$ rather impairs the accuracy of the calculation, and that there is the appropriate range of values for $\gl,\gr$ that do not affect the correct solution of the linear equation.
Furthermore, $N$ and $n$ has the different effect on the accuracy of the calculation, even when the size of the minimization problem is the same.

The formulation method of the CSM also has the significant impact on 
the accuracy.
In particular,
the formulation based on the integration of the residuals is found to be more accurate than the tau method even for a larger expansion order $N$.
Furthermore, calculations with the linear least squares method could also be used in the Fourier spectrum method to compute two-dimensional problems.
However, while calculations with simulated annealing were mostly accurate, those with quantum annealing were still farther from the exact solution.

The differences in computational accuracy due to the parameters and formulation methods suggest that the accuracy depends on not only the computational scale of the minimization problem but also the condition number of the linear equation.
If the resolution of the solution is desired to be high, $N$ and $n$ must be large.
However, the accurate solution may be calculated  with quantum annealing by reducing the condition number by devising other parameters or formulation methods.

In this study, the calculations used the advection-diffusion equation, which is a linear differential equation with no pressure effect, for simplicity.
Despite this simple setting, the present results have revealed a number of challenges
to overcome toward the practical use of quantum annealers for fluid flow simulations.
Although adopting the CSM is found to increase the accuracy as compared to FDM-based annealing, the deterioration in accuracy with the expansion order is still crucial.
It suggests a need of some different formulations which can suppress the condition number of matrix.
Also, the present pessimistic results on the actual quantum annealer also suggests a need of some different algorithms, which do not rely on the assumption of fully-connected network.
As these problems have been elucidated in the present study, the next step toward the practical use of quantum annealers for the next 40 years of fluid dynamics research is to develop such alternative formulations and algorithms more suited to the structure and characteristics of the actual quantum annealers, which are also expected to improve significantly.

\section*{Acknowledgments}
The computer environments for the quantum annealer and its simulator used in the present study were provided by Fixsters Amplify and D-Wave Quantum Inc. 
The authors are grateful to Dr.~Yuki Minamoto (Fixsters Amplify) for the initial instructions.

\appendix
\section{The three formulations for the Chebyshev spectral method}
Here we briefly explain the
three 
standard
formulations for the Chebyshev spectral method, i.e.,
 Galerkin, tau, and collocation,
by taking
the case of $N=3$
as an example.
For more detail, readers are referred to Fornberg (1998).

\bd{Galerkin method}
In the Chebyshev--Galerkin method, the weight function $\psi_k$ is defined as
\begin{eqnarray}
    \psi_k(\xi)=\left\{\matrix{%
        C_+T_k(\xi)+C_-T_1(\xi), & k:\mathrm{even} \cr
        C_-T_k(\xi)+C_+T_0(\xi), & k:\mathrm{odd} \cr
    }\right.,\label{eq:Galpsi}\\
    C_+\equiv\frac{\pl+\pr}{2},\ C_-\equiv\frac{-\pl+\pr}{2},
\end{eqnarray}
such that 
 $\psi_k$
satisfies
the same boundary conditions as the 
solution
$\phi$.
The constraint is then written in the form of \Eref{eq:OD} with
\begin{eqnarray}
    \fl\eqalign{
    \boldsymbol{A}&=\left[\matrix{%
        0 & (\Delta t)^{-1} C_- & 8uC_-+(\Delta t)^{-1} C_+ & -96\alpha C_-+12uC_+ \cr
        2(\Delta t)^{-1}C_+ & 4C_+u & -32\alpha C_+ & 12uC_++(\Delta t)^{-1} C_- \cr
    }\right],\\
    \boldsymbol{b}&=\left[\matrix{%
        C_-a_1^{\imath}+C_+a_2^{\imath} \cr
        2C_+a_0^{\imath}+C_-a_3^{\imath} \cr
    }\right].}\label{eq:Gal}
\end{eqnarray}
\Eref{eq:Gal} allows \Eref{eq:OD} to be solved exactly, 
and it does not 
depend
on $\gl$ and $\gr$.

\bd{Tau method}
We use the Chebyshev polynomial $T_k$
for the weight function $\psi_k$.
Since the orthogonality of the Chebyshev polynomial $T_k$ is expressed by \Eref{eq:orth_o_T}, the constraint is written in the form of \Eref{eq:OD} with
\begin{eqnarray}
    \eqalign{
    \boldsymbol{A}&=\left[\matrix{%
       (\Delta t)^{-1} & 2u & -16\alpha & 6u \cr
        0 & (\Delta t)^{-1} & 8u & -96\alpha \cr
    }\right],\\
    \boldsymbol{b}&=\left[\matrix{%
        (\Delta t)^{-1}a_0^{\imath} \cr
        (\Delta t)^{-1}a_1^{\imath} \cr
    }\right].}\label{eq:Tau}
\end{eqnarray}
In fact,
\Eref{eq:Tau} is the truncation of \Eref{eq:ChebAb}.
Implementation of the tau method is 
easier than the Galerkin method because there is no need to define weight functions according to the boundary conditions.

\bd{Collocation method}The Dirac delta function $\delta(\xi)$ defined by
\begin{eqnarray}
    \int_{-\infty}^{+\infty}f(\xi)\delta(\xi-c)\mathrm{d}\xi=f(c),\ c\in\mathbb{R}\label{eq:def_delta}
\end{eqnarray}
for every continuous function $f:\mathbb{R}\rightarrow\mathbb{R}$ is used for the weight function as
\begin{eqnarray}
    \psi_k(\xi)=\delta\left(\xi-\cos{\frac{k\pi}{3}}\right),\ k=1,2.\label{eq:Colpsi}
\end{eqnarray}
The collocation method 
imposes a
constraint
such that $R=0$ 
 at the Chebyshev collocation points $\xi=\cos{\left(\pi/3\right)},\cos{\left(2\pi/3\right)}$.
Hence,
\begin{eqnarray}
    \fl\eqalign{
    \boldsymbol{A}&=\left[\matrix{%
        (\Delta t)^{-1}  & 2u+\frac{1}{2}(\Delta t)^{-1} & -16\alpha+4u-\frac{1}{2}(\Delta t)^{-1} & -48\alpha-(\Delta t)^{-1}  \cr
        (\Delta t)^{-1}  & 2u-\frac{1}{2}(\Delta t)^{-1} & -16\alpha-4u-\frac{1}{2}(\Delta t)^{-1} & 48\alpha+(\Delta t)^{-1} \cr
    }\right],\\
    \boldsymbol{b}&=\left[\matrix{%
        (\Delta t)^{-1} a_0^{\imath}+\frac{1}{2}(\Delta t)^{-1} a_1^{\imath}-\frac{1}{2}(\Delta t)^{-1} a_2^{\imath}-(\Delta t)^{-1} a_3^{\imath} \cr
        (\Delta t)^{-1} a_0^{\imath}-\frac{1}{2}(\Delta t)^{-1} a_1^{\imath}-\frac{1}{2}(\Delta t)^{-1} a_2^{\imath}+(\Delta t)^{-1} a_3^{\imath} \cr
    }\right].}\label{eq:Col}
\end{eqnarray}

\section*{Reference}

\end{document}